\documentclass[12pt]{iopart}

\usepackage{amssymb,amsfonts}
\usepackage{algorithmic}
\usepackage{graphicx}
\usepackage{textcomp}
\usepackage{xcolor}

\usepackage{dcolumn}
\usepackage{bm}
\usepackage{upgreek}
\usepackage[numbers]{natbib}

\begin{document}

\title{The role of coupling radius in 1D and 2D SQUID and SQIF arrays} 

\author{Marc A. Gal\'i Labarias, Karl-H. M\"uller and Emma E. Mitchell}

\address{CSIRO Manufacturing, Lindfield, NSW, Australia}
\ead{marc.galilabarias@csiro.au}
\vspace{10pt}
\begin{indented}
\item[]August 2021
\end{indented}


\begin{abstract}
We investigate theoretically the effect of the coupling radius on the transfer function in 1D and 2D SQUID arrays with different number of Josephson junctions in parallel and series at 77 K.
Our results show a plateauing of the array maximum transfer function with the number of junctions in parallel. The plateauing defines the array coupling radius which we show increases with decreasing the normalised impedance of the SQUID loop inductance. The coupling radius is found to be independent of the number of junctions in series.
Finally, we investigate the voltage versus magnetic field response and maximum transfer function of one 1D and two 2D SQIF arrays with different SQUID loop area distributions.

\end{abstract}


\section{Introduction}

Superconducting quantum interference devices (SQUIDs) have been extensively studied for their high magnetic field sensitivity \cite{Tesche1977, Clarke2004}. In order to extend the device capabilities, their robustness and sensitivity, SQUID arrays were proposed \cite{Miller1991}.
The concept of a superconducting quantum interference filter (SQIF) was theoretically investigated by Oppenl\"ander et al. \cite{Oppenlander2000}. SQIF arrays were proposed to improve the response and linearity of SQUID arrays. 
In particular, at $T=0$ K, the linearity of these devices has been extensively studied \cite{Kornev2009a}.

The emergence of high-temperature superconductors (HTS) extended the range of applications for SQUID technologies, e.g. HTS SQUIDs for geophysical exploration \cite{Foley1999a}, or SQIF arrays as magnetometers  \cite{Oppenlander2002, Mitchell2016}. Even though theoretical models for SQUID arrays at 0 K have been developed for many years \cite{Miller1991, Oppenlander2000, Cybart2012}, they failed to accurately model devices operating at high temperatures. 
Due to the increasing interest in HTS devices the need for theoretical models capable to accurately simulate these devices arose.
Recently, a theoretical model for one-dimensional SQUID arrays operating at $T=77$ K that includes fluxoid focusing has been developed by M\"uller and Mitchell \cite{Muller2021}. Furthermore, a model for two-dimensional SQUID arrays at 77 K has been formulated by Gal\'i Labarias et al. \cite{GaliLabarias2021}.

One of the most important attributes of SQUID and SQIF arrays is their flux-to-voltage response. Among other parameters, this response depends on the geometry and dimension of the array.
In particular, the response dependence on the number of Josephson junctions in parallel has been investigated at $T=0$ K for 1D SQUID arrays \cite{Mitchell2019}, 1D SQIF arrays \cite{Kornev2009b, Kornev2011} and recently at $T=77$ K for 2D SQUID arrays \cite{GaliLabarias2021}.

In this paper we present a study of the effect of the coupling radius on the transfer function in one- and two-dimensional SQUID and SQIF arrays operating at high temperatures, $T=77$ K.  
The goal of this work is to further investigate the effective number of junctions in 1D and 2D SQUID arrays depending on the array normalised impedance of the SQUID loops, as well as to analyse the response of 1D and 2D SQIF arrays depending on their SQUID loop areas.

\section{Modelling}
For this work we will use the model introduced by Gal\'i Labarias et al. \cite{GaliLabarias2021} which takes into account all the currents circulating in the array and generated fluxes as well as thermal Johnson noise currents. This model assumes grid-like arrays, i.e. SQUIDs in the same row have the same height and SQUIDs in the same column have the same width.
The current $I_k(t)$ through the k\textsuperscript{th} Josephson junction (JJ) is given by the resistively shunted junction (RSJ) model as

\begin{equation}
    I_k(t) = I_c \sin \varphi_k(t) + \frac{\Phi_0}{2\pi R}\frac{d \varphi_k(t)}{dt} + I^{n}_k(t) ,
    \label{eq:current-phase-eq}
\end{equation}
where $t$ is the time, $I_c$ and $R$ are the critical current and normal resistance of the JJs, and $\Phi_0$ the flux quantum. $\varphi_k(t)$ is the time-dependent gauge-invariant phase difference across the k\textsuperscript{th} junction and $I^n_k(t)$ is the noise current created by the Johnson thermal noise at the k\textsuperscript{th} junction. 

\begin{figure}[h]
    \centering
    \includegraphics[scale=0.35]{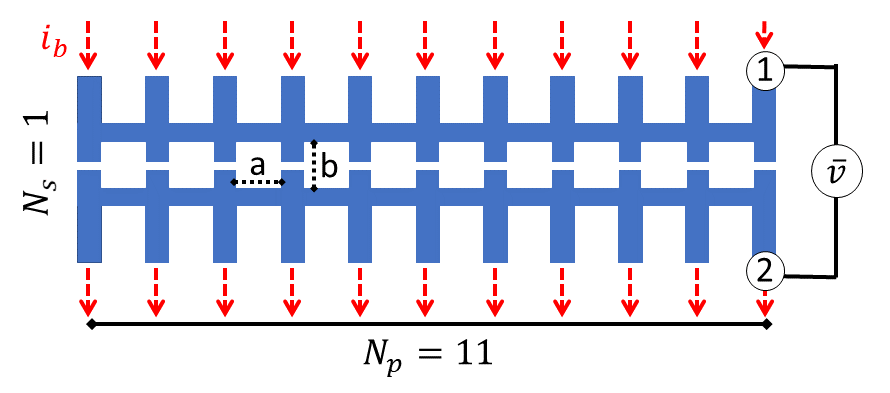}\\
    \includegraphics[scale=0.35]{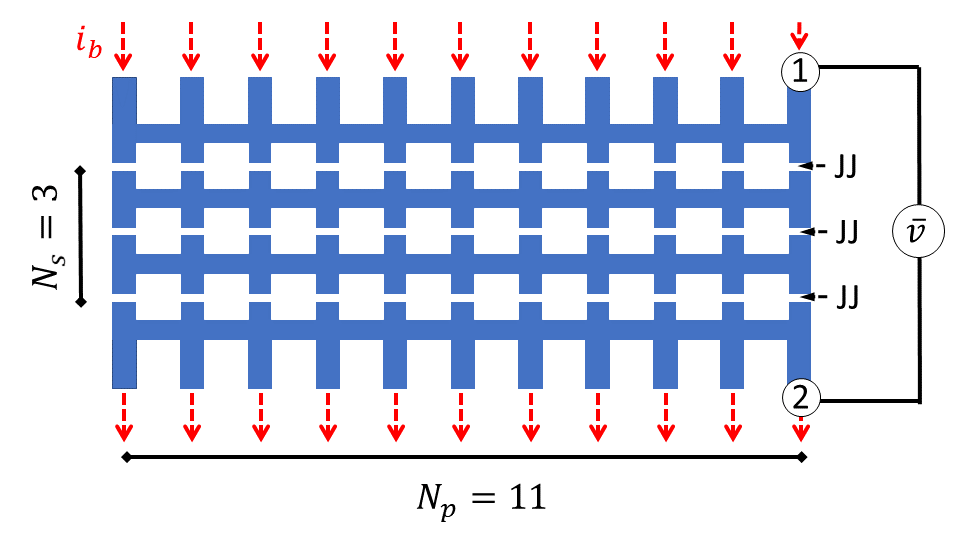}
    \caption{Representation of the square loop SQUID arrays studied. The inner loop height $b$ and width $a$ are equal. The one-dimensional array (top) is a $(1, 11)$-SQUID array, while the two-dimensional array (bottom) is a $(3, 11)$-SQUID array. The gaps indicate the Josephson junctions (JJs).}
    \label{fig:SQUIDarray-diag}
\end{figure}

Using this model one can obtain the flux-to-voltage response for one- and two-dimensional SQUID and SQIF arrays and from the slope of the response one gets the transfer function.
Figure \ref{fig:SQUIDarray-diag} shows the one-dimensional $(1, N_p)$-SQUID array and two-dimensional $(N_s, N_p)$-SQUID array that we investigated, where $N_s$ is the number of JJs connected in series and $N_p$ is the number of JJs connected in parallel. 
For simplicity we only considered SQUID arrays made of square SQUIDs, where the height $b$ and width $a$ of the loops are equal. The bias current is uniformly injected and the voltage is taken between the point 1 and 2 in Fig. \ref{fig:SQUIDarray-diag}.
In this paper we discussed arrays with $N_s=1$ and 3.
Larger arrays have been previously studied in \cite{GaliLabarias2021}.

\subsection{Normalised variables and simulation parameters}
In this paper the following convention is used: The normalised bias current is $i_b=I_b/(N_p I_c)$ where $I_b$ is the total bias current. The normalised applied flux is $\phi_a= \Phi_a/\Phi_0$ with $\Phi_a$ being the applied magnetic flux, and the $RI_c$ normalised time-averaged voltage is $\bar{v}$. 

In this paper we will use $T=77$ K, $I_c=20$ $\upmu$A, film thickness $d=0.2$ $\upmu m$, JJ width $w=2$ $\upmu m$ and
London penetration depth $\lambda$=0.4 $\upmu m$. These values are typical for YBCO devices \cite{Mitchell2016}.
As discussed elsewhere \cite{GaliLabarias2021}, the optimal bias current at 77 K for uniformly biased arrays is $i_b^{opt} \approx 0.75$, therefore in this paper we will be using $i_b=0.75$.

\section{SQUID array maximum transfer function}
\begin{figure}[h]
    \centering
    \includegraphics{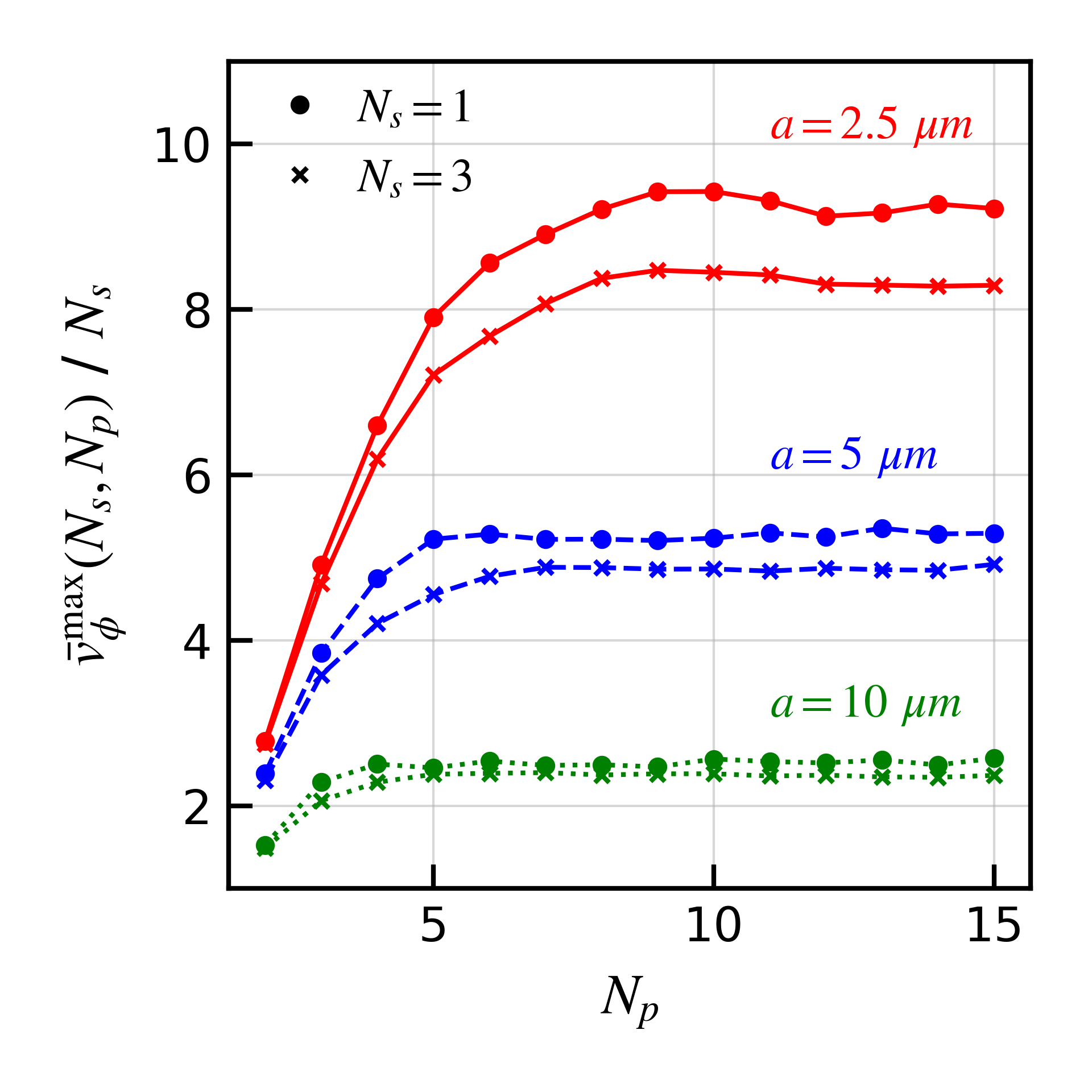}
    \caption{Maximum transfer function versus $N_p$ at 77 K with $i_b=0.75$ for SQUID arrays with different square loop side lengths: in red $a=2.5$ $\upmu m$, in blue $a=5$ $\upmu m$ and in green $a=10$ $\upmu m$.
    Lines with circles picture 1D SQUID arrays and with crosses 2D SQUID arrays ($N_s=3$).}
    \label{fig:dvdphi_vs_Np}
\end{figure}

The flux-to-voltage response of SQUID arrays can be represented by their maximum transfer function $\bar{v}_{\phi}^{\max}=\max( \partial \bar{v}/ \partial \phi_a)$, which maximises at the optimal applied flux $\phi_a^*$. Figure \ref{fig:dvdphi_vs_Np} shows $\bar{v}_{\phi}^{\max}(N_s, N_p)/N_s$ versus $N_p$ for $(1, N_p)$ and $(3, N_p)$-SQUID arrays with different loop sizes: $a=2.5$ $\upmu $m (red), $a=5$ $\upmu $m (blue), $a=10$ $\upmu $m (green). 
This result shows that after a certain number of JJs in parallel $\bar{v}_{\phi}^{\max}$ does not improve further, which can be seen by the plateauing for larger $N_p$. 
The number $N_p^*$ of JJs in parallel beyond which the sensitivity no longer increases is called the coupling radius \cite{Kornev2009b, Kornev2011}.
Figure \ref{fig:dvdphi_vs_Np} further reveals that $\bar{v}_{\phi}^{\max}$ increases with decreasing loop size.

\begin{figure}[htbp]
    \centering
    \includegraphics{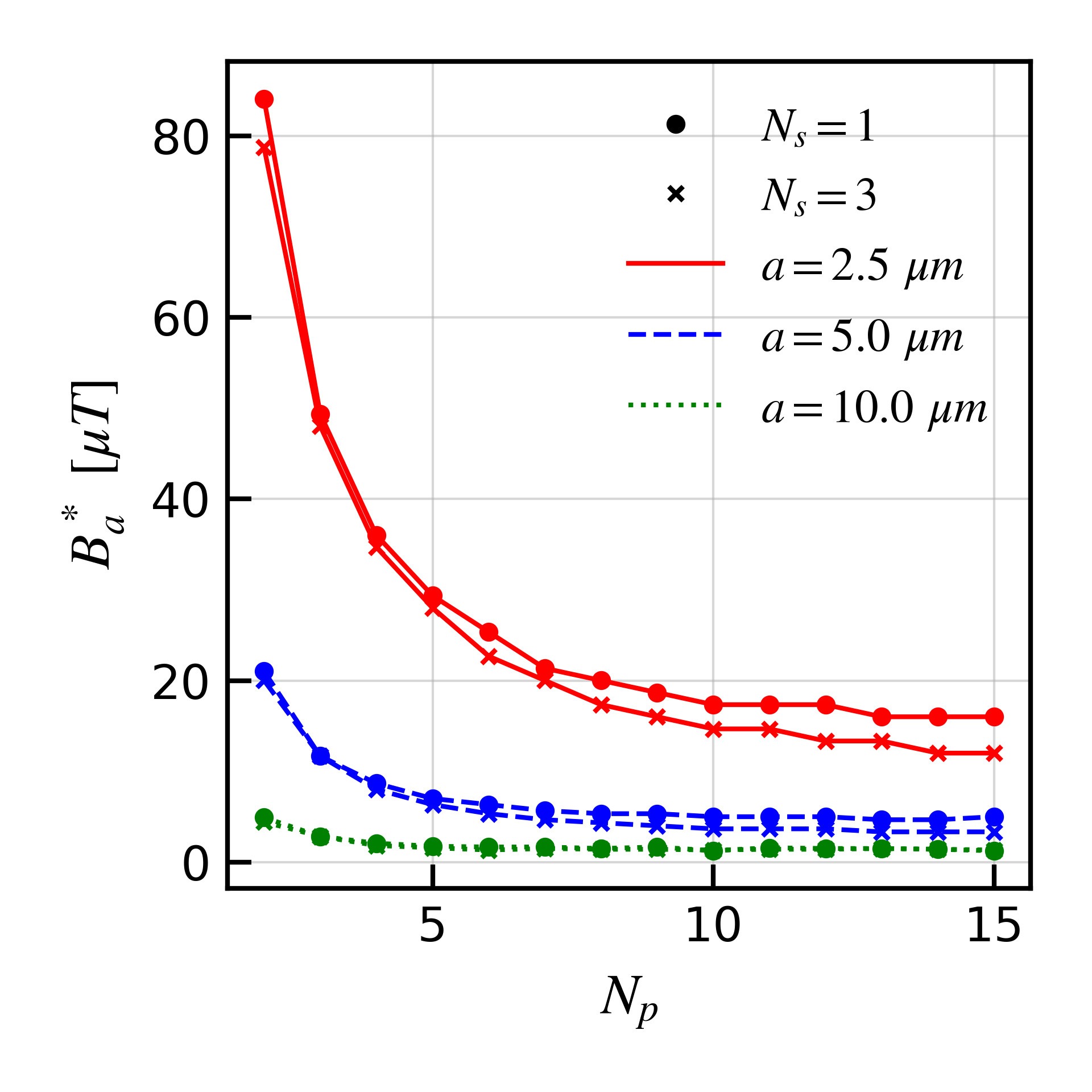}
    \caption{Optimal applied magnetic field $B_a^*$ versus $N_p$ for devices operating at 77 K with $i_b=0.75$ for SQUID arrays with different square loop side lengths: in red $a=2.5$ $\upmu m$, in blue $a=5$ $\upmu m$ and in green $a=10$ $\upmu m$.
    Lines with circles picture 1D SQUID arrays while lines with crosses picture 2D SQUID arrays ($N_s=3$). }
    \label{fig:phi_a-vs-Np}
\end{figure}

Figure \ref{fig:phi_a-vs-Np} shows the optimal applied magnetic field $B_a^*=\Phi_a^*/ a^2$ that maximises the transfer function. From these curves we can see that for the different loop sizes studied, $B_a^*$ is independent of $N_s$ but strongly dependent on $N_p$. This is in agreement with Oppenl\"ander et al. \cite{Oppenlander2000}, where the authors showed that for 1D SQIF arrays operating $T=0$ K the width of the main dip in their response curves decreases with $N_p$, i.e. the transfer function is maximised at smaller $B_a^*$.

\section{Coupling radius of SQUID arrays}
The coupling radius (or interaction radius) $N_p^*$ \cite{Kornev2009b, Kornev2011} defines the number of JJs connected in parallel that can effectively interact with each other. 
For $N_p > N_p^*$ the maximum transfer function of the device does not further improve with $N_p$. As discussed by Kornev et al. \cite{Kornev2009b, Kornev2011} the coupling radius depends on the normalised impedance of the SQUID loops, defined by the product $\omega \cdot l$.
Here $l$ is the normalised SQUID loop inductance (which includes the kinetic inductance term), $l=2\pi I_c L_s/\Phi_0$
and $\omega$ is the normalised intrinsic operating frequency of the array normalised by the characteristic Josephson frequency $\omega_c = 2\pi I_c R/\Phi_0$.
In our case $\omega = \bar{v}^*/N_s$ where $\bar{v}^*$ is the $RI_c$ normalised time-averaged voltage of the array taken at $\phi_a^*$, $i_b^{opt}$ and $N_p=N_p^*$.
 
\begin{figure}
    \centering
    \includegraphics{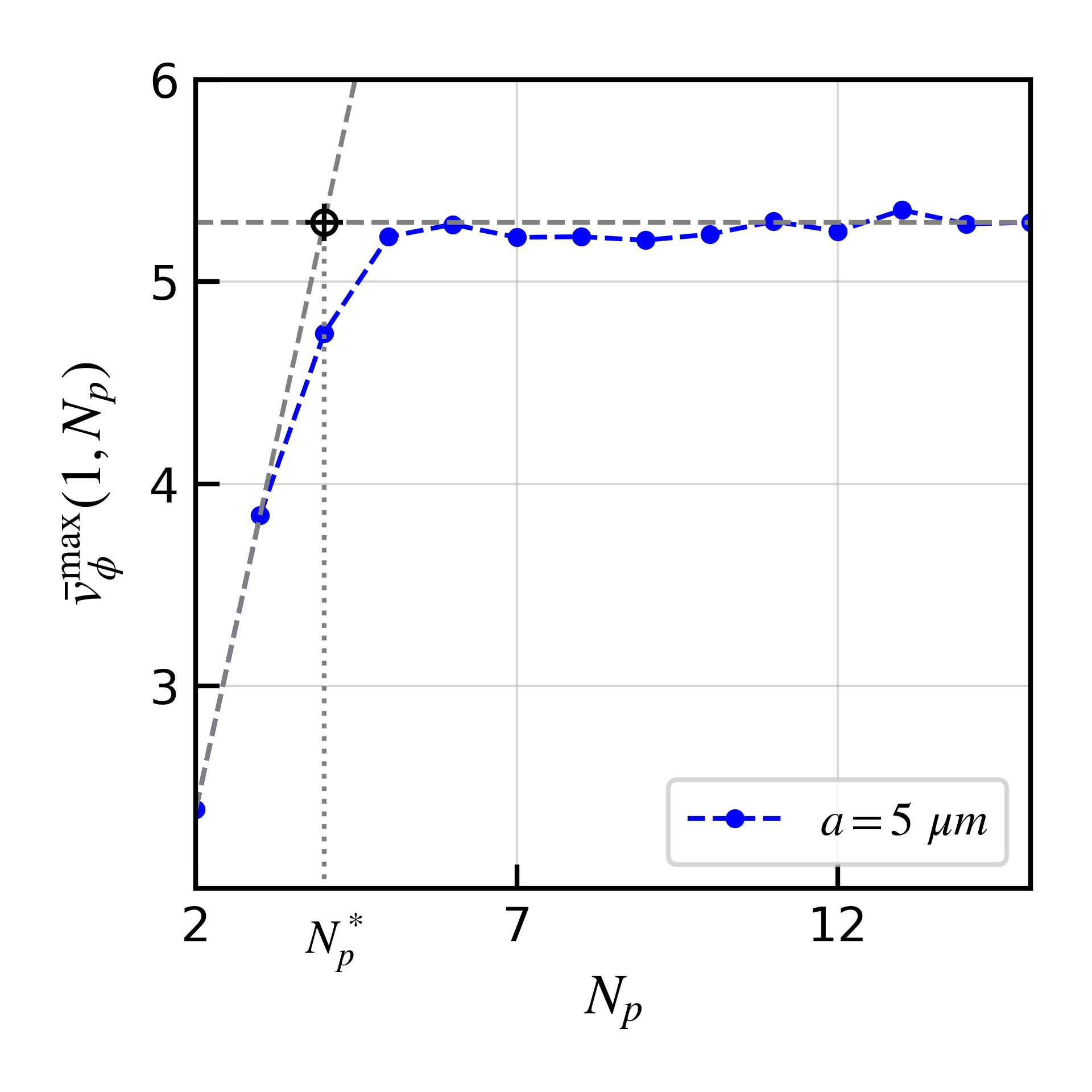}
    \caption{Line method to obtain an approximate value of the coupling radius $N_p^*$. The blue dashed line with circles represents $\bar{v}_{\phi}^{\max}$ vs $N_p$ for $(1, N_p)$-SQUID arrays with $a=5$ $\upmu m$ and $i_b=0.75$ at $T=77$ K.}
    \label{fig:LineApprox}
\end{figure}

 Figure \ref{fig:LineApprox} shows the method used in this paper to obtain an approximate value of $N_p^*$. We define $N_p^*$ as the value on the $N_p$-axis corresponding to the intersection point of the line defined by the points $\bar{v}_{\phi}^{\max}(N_s, 2)$ and $\bar{v}_{\phi}^{\max}(N_s, 3)$, and the constant line defined by  $\bar{v}_{\phi}^{\max}(N_s, \max(N_p))$. 
 
\begin{figure}[h]
    \centering
    \includegraphics{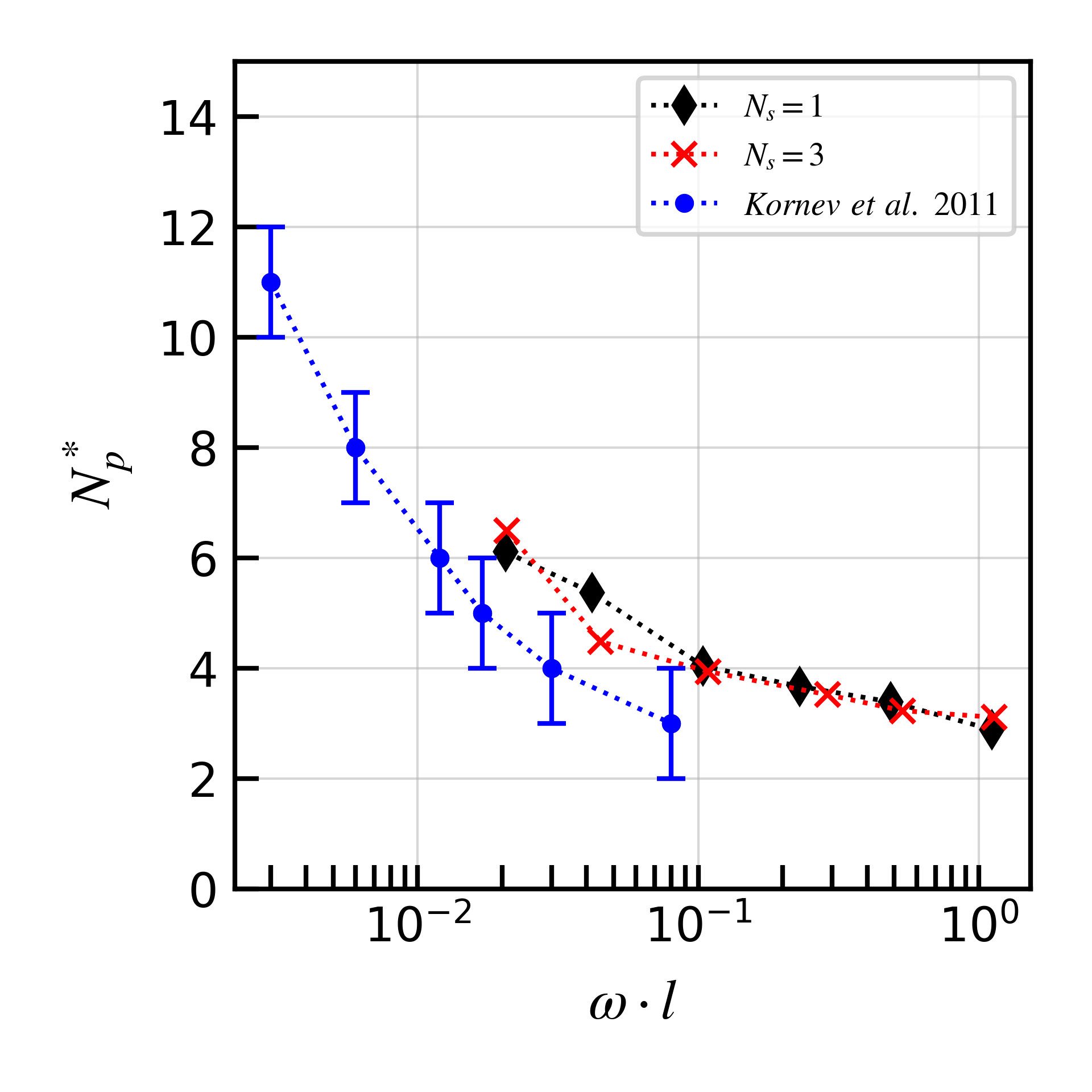}
    \caption{Coupling radius $N_p^*$ versus $\omega \cdot l$ obtained by only varying the array SQUID loop inductance. 
    The black diamonds and red crosses are obtained from $(1, N_p)$- and $(3, N_p)$-SQUID arrays. The square loop side lengths $a$ used are: 0.6 $\upmu m$, 1.25 $\upmu m$, 2.5 $\upmu m$, 5 $\upmu m$, 10 $\upmu m$ and 20 $\upmu m$. For comparison, blue circles show data taken from Kornev et al. \cite{Kornev2011} where $T=0$ K was assumed. }
    \label{fig:coupling_radius}
\end{figure}

Figure \ref{fig:coupling_radius} shows the coupling radius for 1D (black diamonds) and 2D with $N_s=3$ (red crosses) SQUID arrays as well as the data points obtained by Kornev et al. \cite{Kornev2011} for parallel arrays at 0 K. This result reveals that the coupling radius $N_p^*$ is independent of $N_s$ for the loop sizes studied here.
Our $N_p^*$ values are larger than those of Kornev et al. \cite{Kornev2011} which we attribute to the fact that our calculations are done at $T=77$ K allowing us to optimise the bias current $i_b^{opt}$.

\section{SQIF arrays}
\begin{figure}
    \centering
    \includegraphics[scale=0.25]{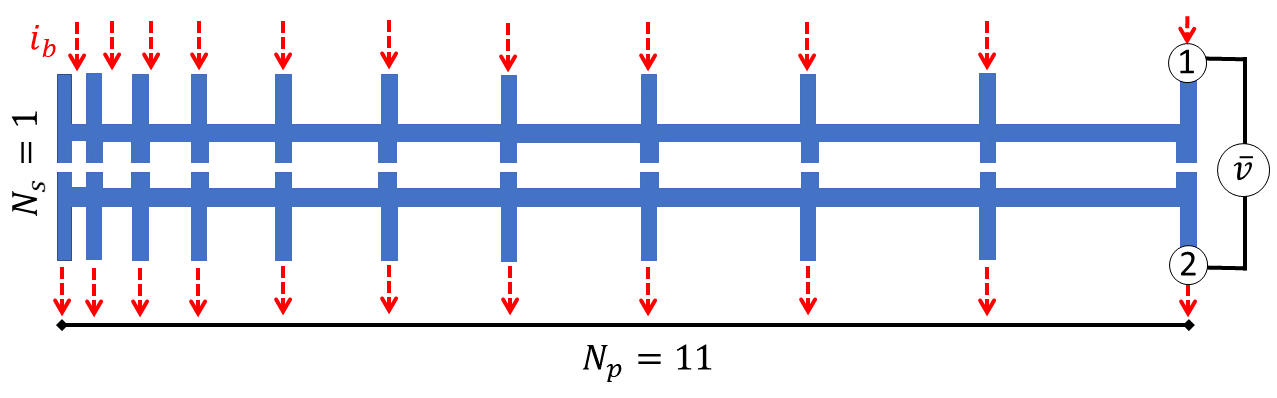}\\
    \includegraphics[scale=0.25]{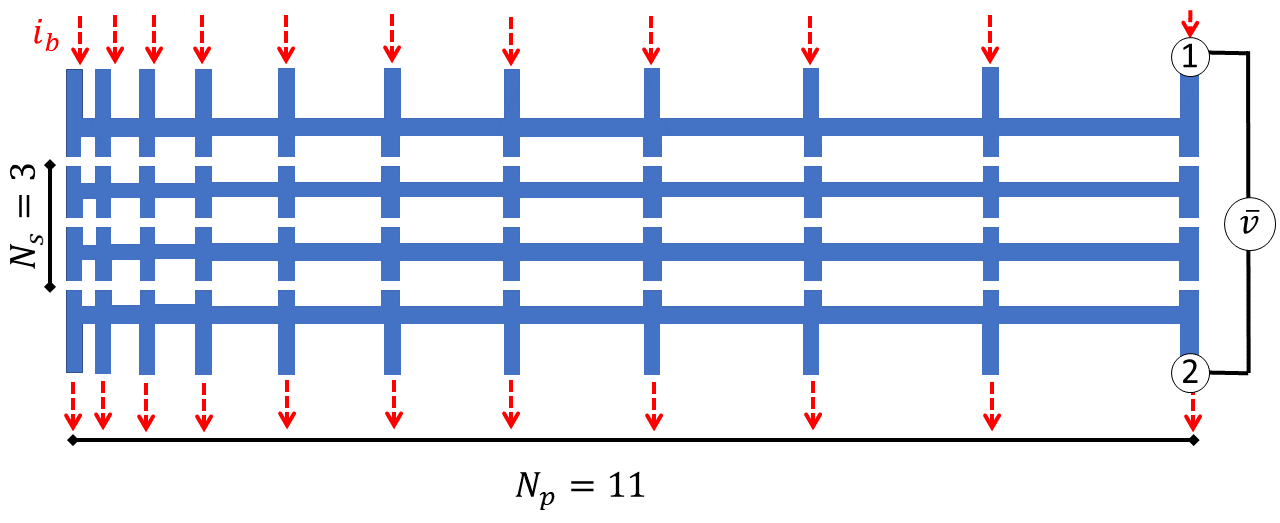}\\
    \includegraphics[scale=0.25]{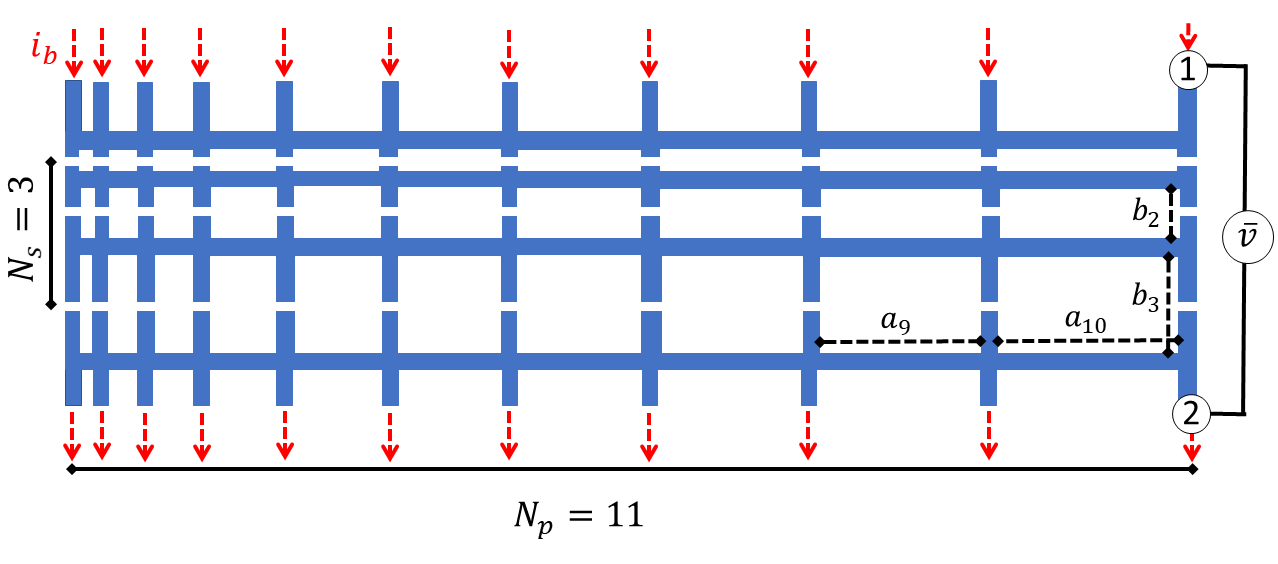}
    \caption{Representation of the SQIF arrays studied. Top, $(1, 11)$-SQIF array with $a_m=(2m-1)a_1$ for a given $a_1$; middle, $(3, 11)$-SQIF array with $b_n=b_1$ and $a_m=(2m-1)a_1$ for a given $a_1$ and $b_1$; bottom, $(3, 11)$-SQIF array with $b_n= (2n -1)b_1$ and $a_m=(2m-1)a_1$ for a given $a_1$ and $b_1$. }
    \label{fig:SQIF-diag}
\end{figure}

A SQIF array consists of SQUIDs with different loop areas, which breaks the SQUID array strict $\Phi_0$-periodicity but preserves the main dip around $B_a=0$. The shape of this main dip is mainly defined by the largest loop area in the SQIF array.
There are an arbitrary number of ways to define the SQIF loop areas. 
In this work we will use the same method as introduced by Oppenl\"ander et al. \cite{Oppenlander2000}, where the loop side lengths follow the progression $a_m=(2m - 1)a_1$ for a given $a_1$. 
It is important to note that the field-to-voltage response of these SQIF arrays shows a periodicity in $B_a$ determined by the smallest area in the array, where the period is $\Delta B_a=\Phi_0/\min(S_i)$, and $S_i$ is the area of the i\textsuperscript{th} loop.
Figure \ref{fig:SQIF-diag} shows a representation of the SQIF arrays that we will investigate.

\begin{figure}[htbp]
    \centering
    \includegraphics[scale=1]{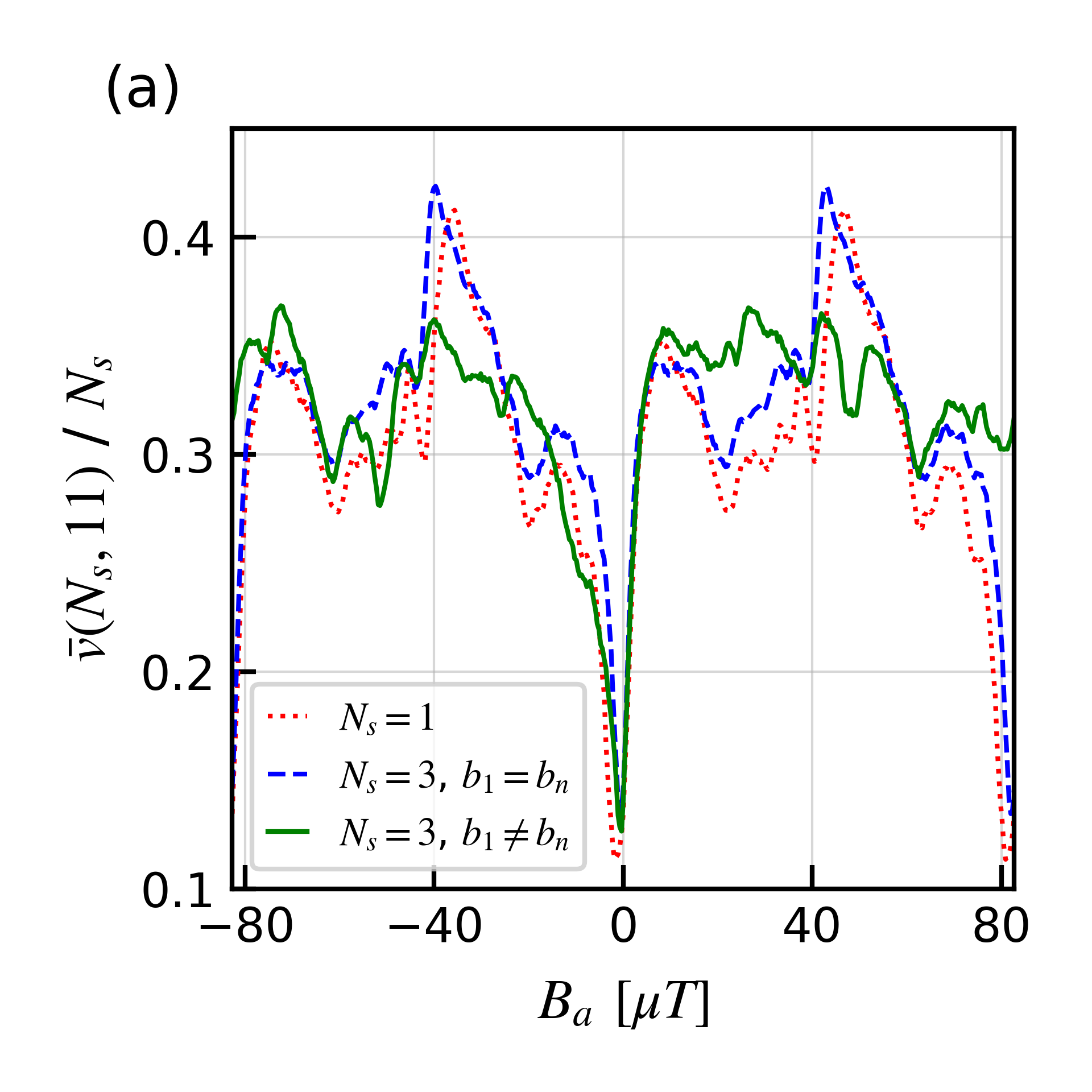}\\
    \includegraphics[scale=1]{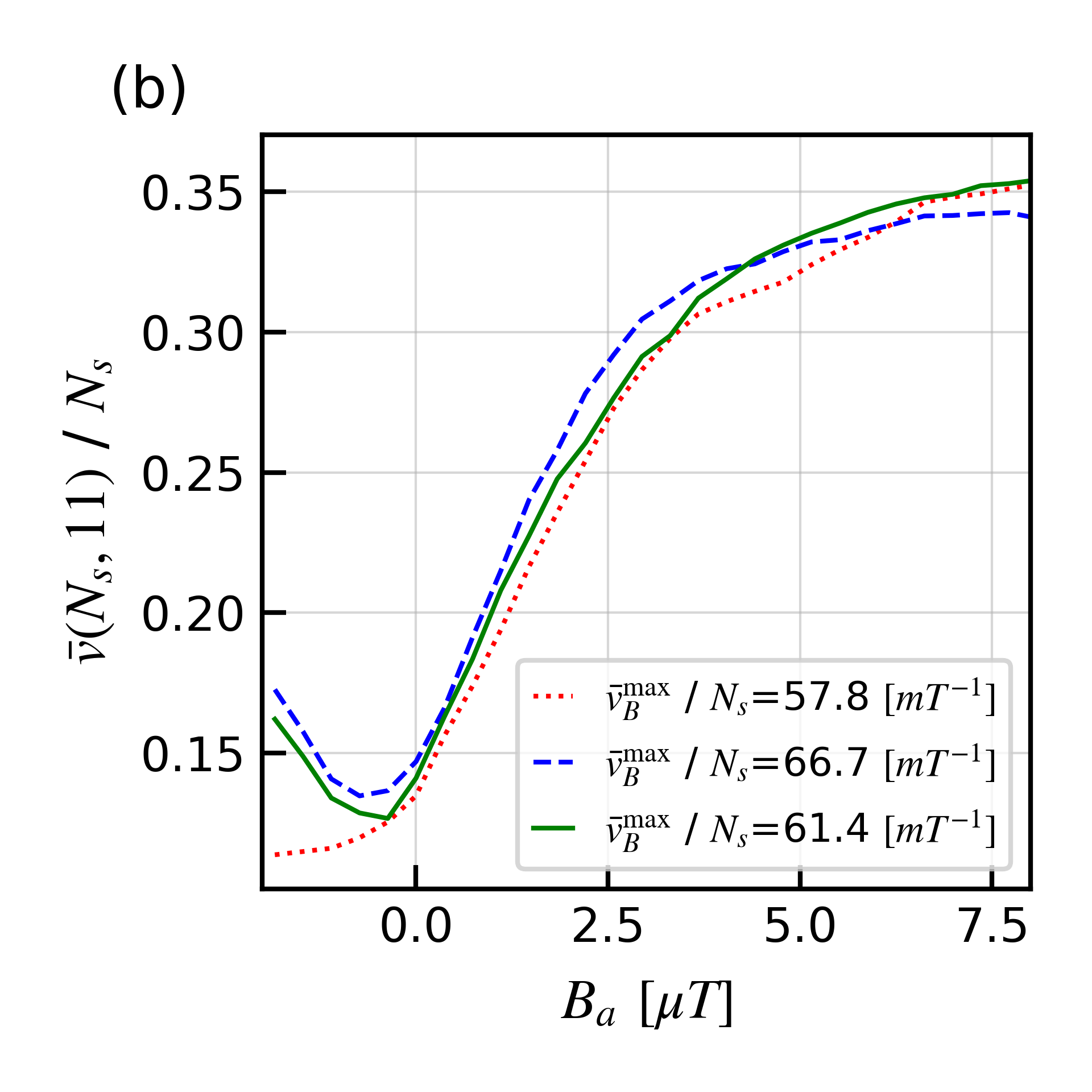}
    \caption{(a) $\bar{v}/N_s$ vs $B_a$ for three different SQIF arrays biased with $i_b=0.75$ and operating at $T=77$ K. Red dotted line, $(1, 11)$-SQIF array with $b=5$ $\upmu m$ and $a_m=(2m-1)a_1$ with $a_1=5$ $\upmu m$; blue dashed line, $(3, 11)$-SQIF array with $b_n=5$ $\upmu m$ and $a_m=(2m-1)a_1$ with $a_1=5$ $\upmu m$; and green solid line $(3, 11)$-SQIF array with $b_n= (2n -1)b_1$ with $b_1=1.66$ $\upmu m$ and $a_m=(2m-1)a_1$ with $a_1=5$ $\upmu m$. (b) Magnified main dip region with the legend showing the $N_s$-normalised maximum transfer function $\bar{v}_B^{\max}/N_s$ of each SQIF response. }
    \label{fig:1D-SQIF}
\end{figure}

Figure \ref{fig:1D-SQIF}(a) shows $\bar{v}(N_s, N_p)/N_s$ versus $B_a$ for different SQIF arrays. The red dotted line describes a $(1, 11)$-SQIF array with $a_1=5$ $\upmu m$; the blue dashed line a $(3, 11)$-SQIF array with $a_1=5$ $\upmu m$ and equal loop heights across the array, i.e. $b_1=b_2=b_3=5$ $\upmu m$. The green solid line depicts a $(3, 11)$-SQIF array with $a_1=5$ $\upmu m$ where the loop heights of each row also increase following the progression $b_n=(2n-1)b_1$.  In order to make a fair comparison with the two other SQIF arrays, we define $b_1=1.66$ $\upmu m$ so that $1/N_s \sum^{N_s}_{n=1} b_n=5$ $\upmu m$.
The $N_s$-normalised responses of the 1D SQIF and the 2D SQIF with equal loop heights seem to be quite similar with only a small decrease in the main dip voltage modulation depth of the 2D SQIF compared to the 1D SQIF. 
Comparing the 1D SQIF response with the 2D SQIF with different loop heights we can see a strong decrease in the secondary oscillations, while the responses are similar at the main dip. 

 Figure \ref{fig:1D-SQIF}(b) compares the three SQIF responses at the main dip region and the legend shows the $N_s$-normalised maximum transfer function, $\bar{v}_B^{\max}/N_s$, of the different SQIF arrays, revealing that the 2D SQIFs have slightly higher $\bar{v}_B^{\max}/N_s$ than the 1D SQIF array.
It is interesting to note that the 2D SQIF with equal loop heights has a slightly larger $\bar{v}_{B_a}^{\max}/N_s$ than the SQIF array with different $b_n$, even though the average loop areas of both arrays are equal.

Our calculations show that the behaviour of the coupling radius $N_p^*$ versus $\omega \cdot \langle l \rangle$ of SQIF arrays is very similar to $N_p^*$ versus $\omega \cdot l $ shown for SQUID arrays in Fig. \ref{fig:coupling_radius}. Here $\langle l \rangle$ is the averaged normalised SQUID loop inductance.

\section{Summary}
In this paper we have studied the effect of the coupling radius on the maximum transfer function in one- and two-dimensional SQUID arrays, with $N_s=1$ and 3, depending on $N_p$ and the SQUID loop size. We have shown that the maximum transfer function plateaus with $N_p$ for all the arrays studied. Our results also showed that $B_a^*$ is independent of $N_s$ and rapidly decreases with $N_p$, i.e. the transfer function is maximised at smaller $B_a^*$ with increasing $N_p$.

To analyse the plateauing of the SQUID array maximum transfer function, we have presented an approximation method to obtain the coupling radius of these arrays. Our results showed that $N_p^*$ depends on the normalised impedance of the SQUID loop inductance, while $N_p^*$ seems to be independent of $N_s$. The decrease of $N_p^*$ with $\omega \cdot l$ at 77 K is similar to results at $T=0$ K reported previously in the literature.

Finally, we investigated the $\bar{v}/N_s$ versus $B_a$ response of one 1D and two 2D SQIF arrays with different SQUID loop area disorder. Our results showed a slight improvement of the $N_s$-normalised maximum transfer function for the 2D SQIF arrays, as well as a decrease in the amplitude of the SQIF secondary oscillations.

The results presented in this paper clarify the concept of the coupling radius for 1D and 2D arrays operating at 77 K. Also, our study on the SQUID loop area disorder of 2D SQIF arrays shows that there is room to further optimise SQIF arrays.

\bibliographystyle{iopart-num} 
\bibliography{EUCAS21}

\end{document}